# WHY BCIS WORK POORLY WITH THE PATIENTS WHO NEED THEM THE MOST?

P. Séguin[1,2,3], E. Maby[1,2], J. Mattout[1,2]

[1]Lyon Neuroscience Research Center, CRNL; INSERM, U1028; CNRS, UMR5292;
Brain Dynamics and Cognition Team, Lyon, F-69000, France
[2]University Lyon 1, Lyon, F-69000, France
[3]Department of Physical Medicine and Rehabilitation, Faculty of Medicine, University Jean Monnet
of Saint Etienne, F-42023 Saint-Etienne, France

E-mail: jeremie.mattout@inserm.fr

ABSTRACT: A major objective of Brain-Computer interfaces (BCI) is to restore communication and control in patients with severe motor impairments, like people with Locked-in syndrome. These patients are left only with limited eye and eyelid movements. However, they do not benefit from efficient BCI solutions, yet. Different signals can be used as commands for non-invasive BCI: mu and beta rhythm desynchronization, evoked potentials and slow cortical potentials. Whatever the signal, clinical studies show a dramatic loss of performance in severely impaired patients compared to healthy subjects. Interestingly, the control principle is always the same, namely the replacement of an impossible (overt) movement by a (covert) attentional command. Drawing from the premotor theory of attention, from neuroimaging findings about the functional anatomy of spatial attention, from clinical observations and from recent computational accounts of attention for both action and perception, we explore the hypothesis that these patients undergo negative plasticity that extends their impairment from overt to covert attentional processes.

## INTRODUCTION

Since its early days, brain computer interfaces (BCI) aim at improving autonomy for people with paralysis. However and despite numerous efforts and improvements, BCI are still not used in clinical routine [1], [2]. In this article, after a brief overview of non-invasive BCI for communication with patients with severe motor impairment, we show how empirical and theoretical results in cognitive neuroscience could help explain the failure of technological transfer that our field is facing.

## THE TARGETED CLINICAL POPULATION

The « classical » locked-in syndrome (LIS) is caused by a lesion of the brainstem, most of the time due to a stroke [3], [4]. The patient is totally paralyzed. Only the vertical eye movements and blinks are spared and allow him to communicate. This condition can also be encountered in other pathologies, like in amyotrophic lateral sclerosis (ALS), a neurodegenerative disease of the motor neurons, at the late stage. Different forms of this disease exist, some affect more the lower motor neurons (progressive muscular atrophy), cortico-spinal motor neurons (primary lateral sclerosis), brainstem motor neurons (bulbar ALS) or cortico-frontobulbar motor neurons (pseudobulbar palsy) [5]. In all these forms, oculomotor muscles are usually preserved, except in the extremely late stages.

## EEG-BASED BCI

To control non-invasive BCI, users have to learn unusual means of interactions. As we will highlight in this article, the signals used as commands for BCI all rely on some form of (covert) selective attentional process.

*Synchronous BCI:*
  *P300:* The most exploited neurophysiological marker in this context is arguably the P300, a positive wave observed in response to external stimuli (about 300ms later), if rare and relevant to the user, surrounded by more frequent but irrelevant stimulations (distractors). If the "P300-speller" mainly uses the visual modality, the P300 wave can also be observed after auditory or tactile stimuli. For selection, the patient has to actively attend the expected target stimulus.

  *Steady-state visual evoked potential (SSVEP):* SSVEP are characterized by a rhythmic neural responses to a flickering stimuli at the frequency of the flicker and its harmonics [6]. BCI protocols use several targets with different flickering frequencies. The amplitude of the SSVEP is increased when the subject pay attention to the target, which allows to detect which stimulus the subject is attending.

*Asynchronous BCI:*
  *Motor imagery:* The imagination of a movement produces a brain activity globally similar to that observed during the actual production of that same movement [7]. A desynchronization (i.e. a drop in EEG signal strength) in the mu and beta frequency bands occurs during the preparation and during the execution or the imagination of a movement. It is measured next to the primary motor cortex, mainly on the contralateral side of the movement. I If the number of commands remains limited, many publications have shown, in healthy volunteers, the ability to control a cursor on a screen. Studies in patients are fewer but the feasibility has been demonstrated [8]. However, these interfaces require a long training, offer limited performance and above all a large number of healthy volunteers are unable to produce the desired signals, to the point of "illiteracy in BCI" for nearly 30% of users [9].

*Slow cortical potentials:* These potentials result in a slow drift of the signal that reflects a change in cortical excitability. A depolarization of a large neuron assembly corresponds to greater excitability, whereas a hyperpolarization corresponds to a greater inhibition. These signals were among the first used in patients. Several healthy subjects and patients managed to communicate using a BCI exploiting these potentials, but at the cost of several months of training and with a writing speed of the order of one character per minute, well below the means for alternative communication when these are accessible [10], [11].

BCI PERFORMANCE IN PATIENTS WITH SEVERE MOTOR IMPAIRMENT

The communication with LIS patients through non-invasive BCI is often put forward by the press when the publication of encouraging but mitigated results is released. Truth is that BCI has been used routinely by patients only in very rare cases [12]. Researchers and clinicians point to the lack of usability of current systems that are still difficult to use on a daily basis without the help of experts [13], [14].

However, looking more closely at the results of the few studies already conducted in patients, and despite their heterogeneous conclusions, another observation is necessary. These interfaces work well, on average, in severely disabled patients who, due to the residual muscle activity they still enjoy, could use another type of interface (lighter and more efficient like a gaze tracking system). Conversely, in patients who most need these BCI because they have more or almost no means of communication, i.e. with abnormal oculomotor movements or in a complete LIS (CLIS, no residual movements, including oculomotor ones), possibilities of restoring communication remain controversial.

Most studies tend to show that brain computer interfaces work poorly with CLIS patients [2], [15]–[17]. Even in LIS, some patients do not achieve a sufficient accuracy. For example a recent study revealed that amongst 37 persons at a late stage of ALS, 9 (24%) could not control a P300-BCI [12].

Two studies published in 2017 claimed that a communication was restored with CLIS patients [18], [19], but some of the methodological aspects of these studies remain unclear, and their results are debated in the BCI community[20].

A longitudinal study, where a patient could learn to use a BCI at the LIS stage and then maintain some control when becoming a CLIS, was more convincing. This study included three subjects with ALS at the LIS state, and one became a CLIS over the 27 months follow up [21].

This patient, a 37-year-old woman who had ALS for 6 years, was asked to control a binary SSVEP-BCI by either focusing on one LED, or ignoring it. From month three onwards, the electro-oculogram could not show any difference between the attended and ignored conditions (confirming the CLIS). However, she performed BCI control with 79% accuracy online, with accuracies above the confidence limits in 18 out of the 27 months and in 27 out of 40 sessions.

This case of communication in CLIS remains an exception, and the reliability of communication happened to be fluctuant.

Some researchers raised the question of a possible "extinction of goal directed thinking" that would accompany the occurrence of a total handicap [16]. This concept was refined later by the same team in terms of "ideomotor silence", which would lead to a loss of voluntary responses and operant learning in long-term paralysis of human patients [23], [24].

WORKING HYPOTHESIS

Severe motor impairment come with and/or yield cognitive deficits such that handling a BCI is made difficult if not impossible. In particular, we postulate that severe motor impairments lead to altered spatial and/or temporal selective attention abilities.

SUPPORTING EVIDENCE

*Persons with severe motor disabilities often present with cognitive impairments*

The neurological examination of these patients is limited by the lack of communication. Some adaptations of the usual technics were necessary to overcome this impairment [25] and to have a better overview of their neurological status.

*ALS*

*Sensory functions:* In ALS patients, the oculomotor system was classically considered as spared [26], [27], but the development of life support systems, exceeding the natural course of the disease, led to observation that oculomotor impairments can be present at the late stage [28]. These impairments can be both peripheral (nuclear) and supranuclear. It is noticeable that the oculomotor system is crucial to maintain interaction with the world. An earlier onset of oculomotor troubles in the ALS has a pejorative effect: it is correlated with a higher probability to progress from the locked-in state to CLIS [29], [30]. ALS subjects present other sensory abnormalities [31], [32].

*Neuropsychological assessment:* 30% of ALS patients (without fulfilling the criteria for dementia) show deficits in fluency, language, social cognition, executive functions and verbal memory. Amongst all the function tested, only visuoperceptive functions were preserved [33].

*Electrophysiological assessment:*
A longitudinal observation in electrocorticography was conducted in a patient with ALS, around the time when he lost all muscle control [22]. Whereas the P300 wave was still observable at the time of this dramatic evolution; three months later, it was no longer detectable.
A recent MEG study showed that the late stage of ALS is associated with higher connectivity in all frequency bands, more scale-free and disassortative brain networks [34]. The auditory event-related potentials are different than in control subjects, namely the location and amplitude of the late positivity, the amplitude of the early negativity (N200), the latency of the late negativity [35] and the mismatch negativity (related to automatic change detection) [36].

*LIS*
*Sensory functions:* neuro-ophtalmological evaluations suggest visual impairment in all locked-in syndrome subjects [37]. This is a multifactorial impairment including binocular diplopia or oscillopsia, refractive errors, dry eye syndrome, keratitis or visual field defect. Tactile perception is usually spared in LIS patients, as the typical cases are due to bilateral ventral pontine lesions, which usually preserve the posterior part of the brainstem that mediates sensory afferences [3]. However, studies assessing the ability of the patients to discriminate the precise location of a stimulus are missing.

*Neuropsychological assessment:* First studies reported intact cognitive abilities in LIS with classical neuropsychological assessments [38]–[41]. More recent studies [42],[43] revealed disparity between patients. Study [42] revealed a decreased short-term memory in 4 out of 10 persons with LIS, and an alteration of executive functions in 2 out of 10. The authors assumed that these impairments could be related to additional cortical or thalamic structural brain lesions [42]Another study highlighted impairments in auditory recognition (associative level), oral comprehension of complex sentences, delayed visuospatial memory, mental calculation and problem solving, compared to matched healthy subjects [43].
Studies testing specifically embodied cognition hypothesis found some abnormalities. LIS subjects have severe troubles to mentally manipulate hand images (a task that imply a mental manipulation of the subject's own hand), whereas they are still able to mentally manipulate images [44]. Contrary to healthy subjects, having the presentation of the hand in the correct physiological side (e.g. left hand on the left) doesn't help them to solve the task. This failure of action simulation was interpreted as a defect of embodied cognition. This points to aspects that are usually not tested in these patients, namely modifications of their peripersonal space.

*Neurophysiological and anatomical data* obtained in LIS reveal some abnormalities that are not yet well explained. Cortical neuronal synchronization mechanisms in the resting state condition are altered [45]. Lugo et al, 2016, report an absence of auditory P300 in four LIS subjects amongst seven tested in a passive condition where subjects were asked to listen only. In an active condition, where the subjects were asked to count the deviants, there were still two LIS subjects out of seven who did not present a P300 [46].
Finally, an anatomical MRI study showed a selective cortical volume loss in patients, in regions that could be linked to an alteration of the mirror neuron system [47], further supporting the idea that embodied simulation process is altered in these patients.

*Chronic complete SCI*
There are contradictory results between behavioral and electrophysiological assessments in SCI subjects.

*Behavioral assessment:* Studies report that patients experienced movement duration and sensation of effort as normal subjects do [48].

*Electrophysiological assessment:* There are abnormal event-related potentials [49] and patterns of cortical activation during motor imagery tasks [50]. In 2013, Lazzaro et al showed some modifications of the auditory P200 latency (earlier than control) and a diminished bilateral posterior P300 amplitude. This was correlated behaviorally with increased false-positive errors and greater variability of response time in the SCI group [51]. These observations were interpreted as a defect of inhibitory functions in both early perceptual encoding processes and later executive functioning that engages contextual and memory-updating operations.

*Impaired motor execution alters motor planning and action selection*
Studies with induced limb immobilization on healthy subjects [52] show that limb nonuse may affect both motor execution and motor imagery performances. These changes happen within days and even hours of immobilization.

*Impaired motor preparation leads to impaired (covert) attention*
Rizzolatti et al. postulated in 1987 that spatial attention is the consequence of activation of the premotor system. They proposed this framework based on the observation of the oculomotor system, especially covert attention. Covert attention is an attentional effort, without any concomitant movement (except involuntary micro-saccades). They assumed a strict link between the covert orienting of attention and the programming of explicit ocular movements. The fact that oculomotor mechanisms and visuospatial attention share similar cortical networks was confirmed by other studies [53]–[55].
Interestingly, motor preparation is so much related to covert attention, that when eyes cannot move to a location, also the attention cannot move [56], [57].
Furthermore, in healthy subjects, when movements are preceded by an invalid motor cue (Posner like paradigm in the motor domain), an increase in reaction times is observed [58], suggesting that motor preparation is a top-down attentional process and that this biasing is mediated at low level in the motor hierarchy.

Despite criticisms of some aspects of the premotor theory of attention, Smith and Schenk proposed in 2012 that a limited version of this theory is still valuable, namely the tight link between exogenous attention and motor preparation [59]. This could be sufficient to impact the P300 marker, which is simultaneously associated with attentional, exogenous and endogenous processes and action selection processes [60], [61].

This is confirmed by some BCI studies addressing the problem of covert attention. Be it in healthy subjects [62] or patients with oculomotor impairments [63], it was observed that BCI performance are much higher or involve much less mental workload when the gaze can be directed toward the target.

To overcome these visual impairments, some auditory BCI were proposed. Surprisingly, in the few clinical studies with auditory BCI, including ours, most of the patients continue to show poor performance compared to healthy subjects [64]. This multi-modal impairment could be explained by the fact that attention would be mediated by a supra-modal network. For instance, orienting attention toward a tactile target triggers an automatic displacement of spatial attention in the visual modality [65]–[67]. Furthermore, a subpopulation of neurons in the frontal eye field (FEF, orienting the gaze) is directly activated by auditory cues. As the FEF is involved in the planning of both covert and overt attentional orienting, an impairment could prevent accurate orientation of attention in the auditory domain when the oculomotor modality is impaired.

*Connectivity changes induced by focal lesions in motor pathways*

The motor networks and the attentional ones are so interdependent that the question remains about the attentional consequences of chronic severe motor impairments. The general concept of connectional diaschisis (opposed to focal diaschisis) was recently proposed to address this question [68]. Connectional diaschisis is defined as the changes of structural and functional connectivity between brain areas distant to a lesion. The authors consider that those changes should be maximal immediately after the insult and then progressively improve, or sometimes even normalize in parallel with clinical function. Diaschisis is only one phenomenon of connectomal changes impact after cerebral injury, amongst other confounding factors like neuroplasticity and vicariation. All of these factors combined could induce important changes in all systems tightly linked with the motor system, including attention networks.

*An action oriented view of cognition*

Our hypothesis is in line with an action-oriented view of cognition, which is currently experiencing a strong revival of interest [69]–[71]. This theory postulates that cognition should not be understood as providing models of the world, but as subsuming action and being grounded in sensorimotor coupling. Thus, a bundle of theoretical as well as empirical and clinical arguments contribute to reinforce the hypothesis that a negative plasticity, generated by the occurrence of an extreme sensorimotor disability including oculomotor ones, could be at the origin of the degradation of some cognitive capacities at the heart of which lie attentional processes. We especially suspect diminished spatial selection abilities with detrimental consequences for the control of non-invasive brain-machine interfaces.

ADDITIONAL FACTORS AND OBSERVATIONS THAT NEED TO BE ACCOUNTED FOR

*Cognitive impairments due to the physiopathological causes of the disease*

*ALS:* ALS is not only a motoneuron disease. There is a continuum between ALS and other degenerative disorders, the most frequent being fronto-temporal dementia. About 20% of ALS patients meet the criteria for fronto-temporal dementia This implies that even in patients that do not fulfil the whole criteria for fronto-temporal dementia, the observed cognitive impairments could be due to some incomplete form of the disease.

*LIS:* BCI studies approach LIS patients as a homogenous group. However, from a clinical point of view, the variety of etiologies should be taken into account when estimating their cognitive status. Indeed, the most common cause of LIS is stroke. Stroke patients have a bad cardio-vascular condition that can lead to multiple small cerebral stroke that impair cognitive status [72]. The problem is similar with traumatic lesions that can induce acute diffuse axonal lesions, frontal and/or occipital cortical lesions. Detecting of these lesions is difficult, as the routine (anatomical) imaging technics lack sensibility in that respect [73]. The finest assessment is done by neuropsychological and ecological assessments in communicating patients with traumatic brain injury, but with LIS patients, this is often not feasible, except in expert teams. Other etiologies like infectious and/or toxic ones share the same risk of multiple direct injuries of the central nervous system.

*Chronic complete SCI:* when the etiology is traumatic, a traumatic brain injury is frequently associated, in at least one third of the cases [74]. Some studies highlighted the cognitive impact of breathing disorders like sleep apnea induced by the tetraplegia in chronic [75]. Sleep apnea concerns up to 91% of these chronic complete SCI patients [76]. More severe chronic apnea was associated with decrease performance in verbal attention and concentration, immediate and short-term memory, cognitive flexibility, internal scanning and working memory.

*Differences in performance due to a bias in patient selection*

Only a few BCI studies addressed the direct comparison between performance of healthy subjects and patients, and even less did control for age, gender and socio-professional status. Other authors noted that the healthy subjects included in the BCI studies are on average younger and more educated than patients, whereas these factors impact cognitive abilities and electrophysiological data. This bias could explain in part the decrease of performance reported in patients.

Besides, the publication bias favors the reporting and diffusion of successful BCI studies. This yields a probable overestimation of patients' performance.

CONCLUSION

Years of BCI attempts to help patients with severe motor impairment have revealed a paradox. These patients have long been regarded as cognitively intact and motivated enough to be the ideal candidates to benefit from neurotechnologies made for communication. However, the performance of BCI in these patients is lower than in healthy controls and no BCI solution is routinely used by these patients who have no other solution yet. This is endorsed by the BCI studies we briefly reviewed. Most importantly, it urges our community to consider the possible factors that beyond imperfect BCI systems and algorithms, may explain this striking observation.

In this short paper, we first question the simplistic view of those patients as being purely motor impaired, with intact cognitive function. It turns out that whatever the etiology, cognitive impairments do come with motor ones.

Precisely, we hypothesize that the attentional processes could be altered in these patients, concomitantly with their motor disorders or because of a negative plasticity resulting from motor disorders. In particular deficits in oculomotricity as selective spatial attention has been shown to be closely related to overt eye movements. This is endorsed by prevalent theories in cognitive neuroscience, namely the premotor theory of attention which is supported by both behavioral and neurophysiological evidence. And this could explain poor BCI performance as most protocols heavily rely on spatial attention.

Interestingly, this view is also endorsed by recent computational theories of the brain that tightly couple perception and action, and namely attention as a gain control process to select visual features and eye movements towards salient spatial locations [77].

We conclude that improving the way we approach those patients requires addressing those questions in all their complexity, by using neuropsychological assessments, by considering the heterogeneity of etiologies and patient's trajectories, by testing patients in a longitudinal and multi-dimensional manner and also by reporting negative BCI results. This might be the only way to design new and efficient BCI solutions.

This work was supported by the following grant from the French government: ANR-17-CE40-0005, Mind Made Clear.